\documentstyle[aps,pra,epsfig]{revtex}
\begin{document}
\tightenlines
\title{Finite Temperature Perturbation Theory for a Spatially
Inhomogeneous Bose-condensed Gas}
\author{P.O. Fedichev and G.V. Shlyapnikov}
\address{{FOM Institute for Atomic and Molecular Physics AMOLF,
Kruislaan 407, 1098 SJ, Amsterdam, The Netherlands}\\
{Russian Research Center Kurchatov Institute, Kurchatov Square,
123182 Moscow, Russia}}
\date{\today}
\maketitle

\begin{abstract}
We develop a finite temperature perturbation theory (beyond the mean field)
for a Bose-condensed gas and calculate 
temperature-dependent damping rates and energy shifts for Bogolyubov 
excitations of any energy.
The theory is generalized for the case of excitations in a spatially
inhomogeneous (trapped) Bose-condensed gas, where we emphasize
the principal importance of inhomogeneity of the condensate density
profile and develop the method of calculating the self-energy functions.
The use of the theory is demonstrated by calculating the damping
rates and energy shifts of low-energy quasiclassical excitations, i.e.
the quasiclassical excitations with energies much smaller than the mean
field interaction between particles.
In this case the boundary region of the condensate plays a crucial
role, and the result for the damping rates and energy shifts is
completely different from that in spatially homogeneous gases.
We also analyze the frequency shifts and damping of sound waves in
cylindrical Bose condensates and discuss the role of damping in the
recent MIT experiment on the sound propagation.
\end{abstract}

\vspace{4mm}
\tightenlines

\section{Introduction}
Recent developments in the physics of ultra-cold gases have
led to the discovery of Bose-Einstein condensation (BEC) in trapped clouds
of alkali atoms \cite{BEC96JILA,BEC96RICE,BEC96MIT} and
stimulated a tremendous boost in theoretical studies of weakly interacting
Bose gases.
As in previous years, these studies rely
on the binary approximation for the interparticle interaction. The latter
is characterized by the 2-body scattering length $a$, which assumes the
presence of a small gaseous parameter $na^3\ll 1$ (n is the gas density).
Especially intensive are the attempts to reach beyond the ordinary mean
field approach and to develop a 
theory which can properly describe the behavior of finite temperature
elementary
excitations of a trapped Bose-condensed gas and in particular, explain the
JILA \cite{JILAshifts} and MIT \cite{MITshifts}
experiments on energy shifts and damping rates
of the excitations.

The commonly used mean field theory (for $a>0$) is based on the Bogolyubov
quasiparticle approach developed originally for a spatially homogeneous
Bose-condensed gas at $T\rightarrow 0$ \cite{Bogolyubov} and employed by
Lee and Yang \cite{LeeYang} (see also \cite{P}) at finite temperatures.
The generalization of the Bogolyubov method for spatially inhomogeneous
systems has been described by De Gennes \cite{DGennes}.
In the case of a Bose-condensed gas it should be completed by the equation
for the wavefunction of the spatially inhomogeneous condensate, derived by
Pitaevskii \cite{Pit} and Gross \cite{Gross}.

For spatially homogeneous gases the theory beyond the mean field approach
was also developed.
Beliaev \cite{Belyaev} constructed the
zero-temperature diagram technique which allows one to find
corrections to the energies of Bogolyubov excitations, proportional to
$(n_0a^3)^{1/2}$, where $n_0$ is the condensate density. The corrections
are provided by the interaction between the excitations (in particular,
through the condensate) and contain
both real (energy shift) and imaginary (damping rate) parts. At $T=0$ the
latter originates from spontaneous decay of a given excitation ($\nu$) to 
two other excitations ($\gamma$ and $\gamma'$), with smaller energies and 
momenta: $\nu\rightarrow\gamma+\gamma'$. 
A universal expression for the chemical
potential in terms of the self-energy functions has been found by Pines and
Hugenholtz \cite{PinesH}.
It should be emphasized that the corrections
proportional to $n_0a^3$ already depend on the contribution of 3-body
interactions and, hence, can not be obtained within the binary approximation.

The Beliaev approach was employed by Popov \cite{PopovBook} at finite
temperatures. In this case
the corrections to the Beliaev self-energies contain infra-red singularities,
i.e. they tend to infinity for momenta $p\rightarrow 0$. This prompted
Popov to make a renormalization of the theory, which links the microscopic 
approach with phenomenological Landau hydrodynamics \cite{LL}. The Popov 
theory eliminates the infra-red singularities and allows one to describe the 
behavior of low-energy excitations (phonons) at temperatures much smaller 
than the mean field interparticle interaction $n_0\tilde U$
($\tilde U=4\pi\hbar^2a/m$, with $m$ being the atom mass).
The damping of phonons in this temperature range is determined by the Beliaev
damping processes and has also been calculated by Hohenberg and Martin \cite{HM}.
A simplified approach within the dielectric formalism was used by Szepfalusy 
and Kondor \cite{SK} for
calculating the damping rates of excitations in the phonon branch of
the spectrum.
They found that at
temperatures $T\gg n_0\tilde U$ the damping rate of a given excitation
($\nu$) originates from the scattering of thermal excitations
($\gamma$ and $\gamma'$) on the excitation $\nu$ through the processes
$\nu+\gamma\leftrightarrow\gamma'$.
Since the characteristic energies of the thermal excitations $\gamma$,
$\gamma'$ turn out to be much larger than the energy of the excitation
$\nu$, this damping channel can be treated as Landau damping.
It should be noted that the damping rates can be simply found by
considering the interaction between the excitations as a small
perturbation and using Fermi's golden rule.
This allows one to properly take into account the Bogolyubov nature of the
thermal excitations.
The damping rates of phonons in a spatially homogeneous
Bose-condensed gas, in particular for the Szepfalusy-Kondor mechanism,
have been calculated in the recent contributions 
\cite{HS,VL,Recent,FSW,Recent1}.

In order to reach beyond the mean field theory at
$T\agt n_0\tilde U$ one should further develop the Popov approach.
One can also proceed along the lines of the Beliaev theory, since any 
physical quantity should be determined by combinations of the Beliaev
self-energies, which do not contain the infrared singularities.
In this paper we choose the latter way and construct the perturbation
theory for a Bose-condensed gas, which allows us to find the next to 
leading order terms (the terms proportional to
$(n_0a^3)^{1/2}$) in the energy spectrum of the elementary excitations.
As in \cite{SK,HS,VL,Recent,FSW,Recent1}, 
we consider the excitations in the so-called collisionless regime, 
where their De Broglie wavelength is much smaller than the mean free path 
of the thermal excitations.

We start with the case of a spatially homogeneous Bose-condensed gas and
find temperature-dependent energy shifts and damping rates for Bogolyubov 
excitations of any energy.
At temperatures $T\gg n_0\tilde U$ the small parameter of the theory
proves to be
\begin{equation}\label{small}
\frac{T}{n_0\tilde U} (n_0 a^3)^{1/2}\ll 1,
\end{equation}
in contrast to $n_0 a^3\ll 1$ for $T=0$. The appearance of the extra factor
($T/n_0\tilde U$) originates from the Bose occupation numbers of thermal
excitations with energies of order $n_0\tilde U$, which are the most 
important in the perturbation theory. 
As shown below, the damping of excitations
with energies $\varepsilon_{\nu}\sim n_0\tilde U$ is determined by both the
Szepfalusy-Kondor ($\nu+\gamma\leftrightarrow \gamma'$) and
Beliaev ($\nu\leftrightarrow \gamma+\gamma'$) processes,
and can no longer be treated as Landau damping.

The theory is generalized for the case of excitations in a spatially
inhomogeneous (trapped) Bose-condensed gas. A new ingredient here is
related to the inhomogeneous density profile of the condensate and the
discrete structure of the excitation spectrum. We develop the method of
calculating the self-energy functions and derive the equations for finding
the wavefunctions and energies of the excitations (generalized
Bogolyubov-De Gennes equations).

The use of the theory is demonstrated by two examples.
The first one concerns quasiclassical low-energy
excitations of a trapped Bose-condensed gas in the Thomas-Fermi regime.
The term "low-energy" assumes that the excitation energy
$\varepsilon_{\nu}$ is much smaller than the mean field interparticle
interaction $n_{0m}\widetilde{U}$ ($n_{0m}$ is the maximum condensate
density), and the quasiclassical character of the excitations requires
the condition $\varepsilon_{\nu}\gg\hbar\omega$, where $\omega$ is the
characteristic trap frequency.
We consider anisotropic harmonic traps, where the discrete structure
of the excitation spectrum is not important (see below and in \cite{FSW}).
On the contrary, the inhomogeneity of
the condensate density profile has a crucial consequence for the damping
rates and energy shifts of quasiclassical low-energy excitations.
The most important turns out to be the boundary region of the
condensate, where $n_0({\bf r})\tilde U\sim \varepsilon_{\nu}$ \cite{FSW}.
Therefore, the result for the damping rates and energy shifts is
completely different from that in spatially homogeneous gases.

Finally, we analyze the frequency shifts and damping of axially propagating
sound waves in cylindrical Bose condensates.
As found, the nature of damping is similar to that in the case of phonons
in spatially homogeneous Bose condensates.
We show that the attenuation of axially propagating sound wave packets in
the recent MIT experiment \cite{LastKetterle} can be well explained as a
consequence of this damping.

\section{General equations}

We consider a weakly interacting Bose-condensed gas confined in an external
potential $V({\bf r})$. The grand canonical Hamiltonian of the gas can be
written as $\hat{H}=\hat{H}_{0}+\hat{H}_{1\text{ }}$, where (hereinafter
$m=\hbar =1$)
\begin{equation}
\hat{H}_{0}=\int d^{3}r\hat{\Psi}^{\dagger }({\bf r})\left( -\frac{\Delta }{2%
}+V({\bf r})-\mu \right) \hat{\Psi}({\bf r}),  \label{Ham0}
\end{equation}
and the term
\begin{equation}
\hat{H}_{1}=\frac{\widetilde{U}}{2}\int d^{3}r\hat{\Psi}^{\dagger }({\bf r})%
\hat{\Psi}^{\dagger }({\bf r})\hat{\Psi}({\bf r})\hat{\Psi}({\bf r}),
\label{Hpert}
\end{equation}
assumes a point interaction between atoms.
The field operator of atoms $\hat{\Psi}({\bf r})$ can be represented as
the sum of the above-condensate part $\hat{\Psi}^{\prime }$ and the condensate
wavefunction $\Psi _{0}\!=\!\langle \hat{\Psi}\rangle $ which is a {\it c}%
-number. As the interparticle interaction $\hat{H}_{1}$ contains both 
terms conserving the number of above-condensate particles and terms
transferring two above-condensate particles to the condensate (or two
condensate particles to the above-condensate part), the diagram technique
should include both the normal Green function ${\cal G}$ and the anomalous 
Green function ${\cal F}$ (see, e.g. \cite{Belyaev}). 

\begin{figure}
\epsfig{file=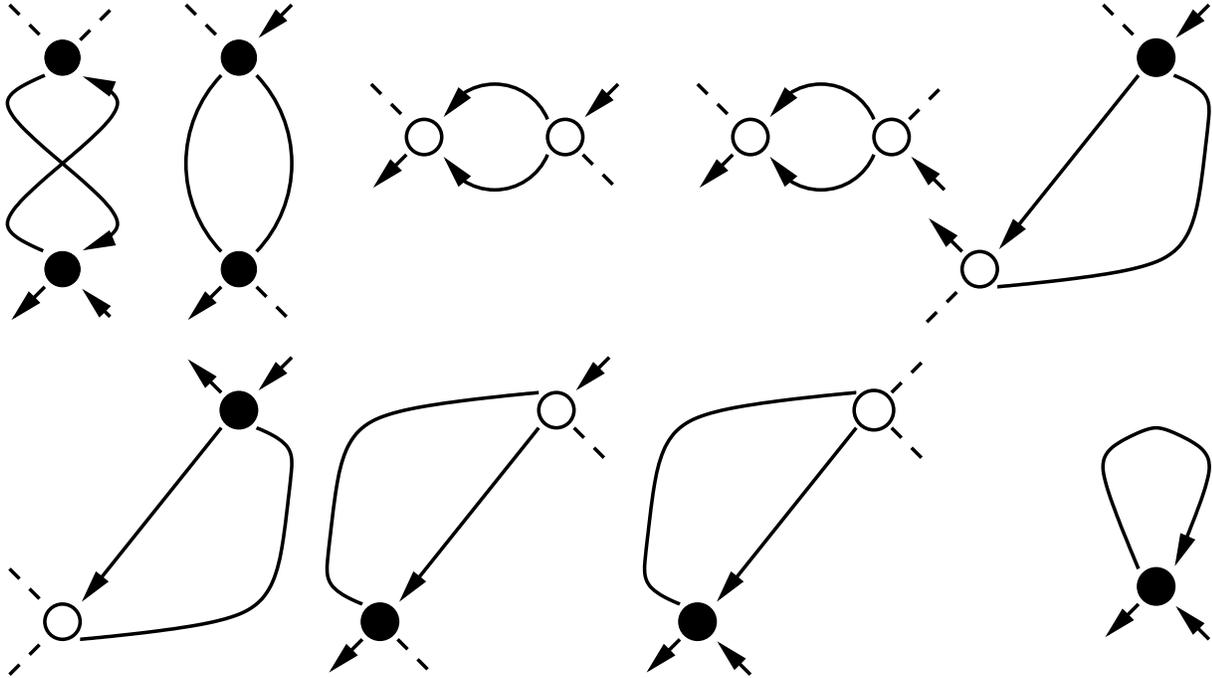, width=0.9\linewidth}
\caption{ The set of diagrams
contributing to the normal self-energy $\Sigma$. Here
a solid line with an arrow represents the normal Green function
${\cal G}$, solid
line without an arrow corresponds to the anomalous Green function
${\cal F}$, white
circle stands for the interaction vertex $\tilde U$ and the black circle
represents a sum of two white circles, one being a direct interaction
and the other an exchange interaction. Dashed lines stand for the condensate
wave function $\sqrt{n_0}$. The self-energy part $\Sigma^+$ can be obtained
by a time-reversal(i.e. the  change $t\rightarrow -t$ and
${\bf p}\rightarrow -{\bf p}$) of the graphs shown above.}
\label{fig1}
\end{figure}

\begin{figure}
\epsfig{file=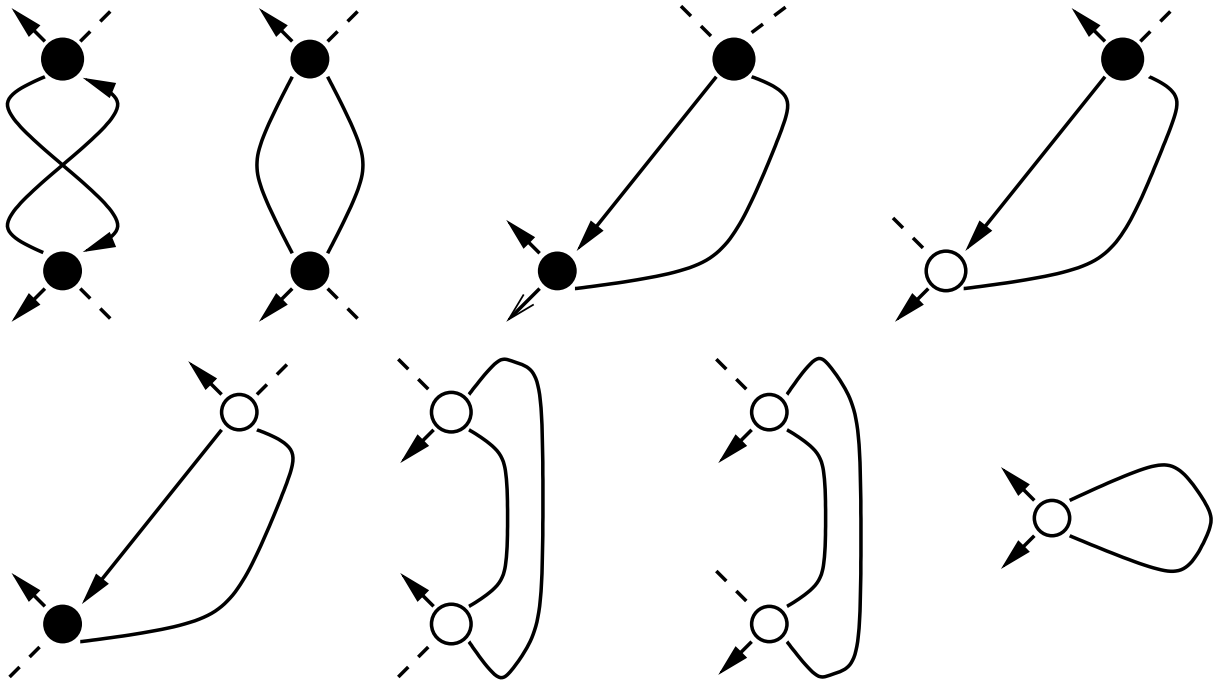, width=0.9\linewidth}
\caption{The set of graphs contributing to the anomalous self-energy
$\Sigma_a$. The notations are the same as for the Fig.1.}
\label{fig2}
\end{figure}

The sums of the
contributions of all irreducible diagrams will be represented by the normal (%
$\Sigma $) and anomalous ($\Sigma _{a}$) self-energies (see Fig.1 and Fig.2).
The former corresponds to the processes conserving the number of
above-condensate particles, and the latter describes absorption (or
emission) of two particles to (out of) the condensate. 
The Green function
and self-energy operators satisfy Beliaev-Dyson equations \cite
{Belyaev,PopovBook}
\begin{eqnarray}
{\cal \,G} &=&G+G\Sigma {\cal G}+G\Sigma _{a}{\cal F},  \label{BOperG} \\
{\cal F} &=&G^{+}\Sigma ^{+}{\cal F}+G^{+}\Sigma _{a}{\cal G},
\label{BOperF}
\end{eqnarray}
where the Green functions $G$ and $G^{+}$ describe forward and 
backward propagation of a particle characterized by the Hamiltonian 
$\hat{H}_{0}$.

We confine ourselves to the case of repulsive interaction between the 
atoms ($a>0 $). To develop the finite temperature perturbation theory for
calculating dynamic properties and finding the excitation spectrum of a
weakly interacting Bose-condensed gas we will use the non-equilibrium
generalization \cite{AGD} of the Matsubara diagram technique. In 
Eqs.~(\ref{BOperG}),(\ref{BOperF}) we perform an analytical continuation 
of the Matsubara frequencies $\zeta _{j}=2\pi Tj$ ($j$ is an integer number)
to the upper half-plane, which corresponds to the replacement $i\zeta
_{j}\rightarrow \varepsilon +i0$. Then, multiplying both sides of Eqs.~(\ref
{BOperG}) and (\ref{BOperF}) by $G^{-1}$ and $(G^{+})^{-1}$, respectively,
we arrive at the system of equations in the frequency-coordinate
representation:
\begin{eqnarray}
\varepsilon {\cal G}(\varepsilon ;{\bf r},{\bf r}^{\prime }) &=&\left[ -%
\frac{\Delta }{2}+V({\bf r})-\mu +\Sigma (\varepsilon )\right] {\cal G}%
(\varepsilon ;{\bf r},{\bf r}^{\prime })+\Sigma _{a}(\varepsilon ){\cal F}%
(\varepsilon ;{\bf r},{\bf r}^{\prime })+\delta ({\bf r}-{\bf r}^{\prime }),
\label{BCoordG} \\
-\varepsilon {\cal F}(\varepsilon ;{\bf r},{\bf r}^{\prime }) &=&\left[ -%
\frac{\Delta }{2}+V({\bf r})-\mu +\Sigma ^{+}(\varepsilon )\right] {\cal F}%
(\varepsilon ;{\bf r},{\bf r}^{\prime })+\Sigma _{a}(\varepsilon ){\cal G}%
(\varepsilon ;{\bf r},{\bf r}^{\prime }).  \label{BCoordF}
\end{eqnarray}
Here the action of the integral self-energy operators on the Green functions
is written in a compact form
$\int d^3r^{\prime\prime}\Sigma(\varepsilon ;{\bf r},
{\bf r}^{\prime\prime}){\cal G}(\varepsilon ;{\bf r}^{\prime\prime},
{\bf r}^{\prime})\equiv
\Sigma(\varepsilon){\cal G}(\varepsilon ;{\bf r},{\bf r}^{\prime})$
(and similar relations for the other combinations).
Eqs.~(\ref{BCoordG}),(\ref{BCoordF}) should be completed by a generalized
Gross-Pitaevskii equation for the condensate wavefunction:
\begin{equation}
\left[ -\frac{\Delta }{2}+V({\bf r})-\mu +(\Sigma -\Sigma
_{a})|_{\varepsilon\rightarrow 0}\right] \Psi _{0}({\bf r})=0  \label{GPeq}
\end{equation}
and by the normalization condition
\[
\int d^{3}r\left( n_{0}({\bf r})+n^{\prime }({\bf r})\right) =N,
\]
where $n_{0}({\bf r})=|\Psi _{0}({\bf r})|^{2}$ is the condensate density, $%
n^{\prime }({\bf r})=\left\langle \Psi ^{\prime \dagger }({\bf r})\Psi
^{\prime }({\bf r})\right\rangle$ is the density of above-condensate
particles, and $N$ the total number of particles in the gas.

Eqs.~(\ref{BCoordG}),(\ref{BCoordF}) can be solved by using the Bogolyubov
transformation for the Green functions:
\begin{eqnarray}
{\cal G}(\varepsilon_{\nu};{\bf r},{\bf r}^{\prime }) & = & u_{\nu}({\bf r})
u_{\nu}^{*}({\bf r}^{\prime })+v_{\nu}({\bf r})v_{\nu}^{*}
({\bf r}^{\prime }), \nonumber \\
{\cal F}(\varepsilon_{\nu};{\bf r},{\bf r}^{\prime }) & = & -u_{\nu}({\bf r})
v_{\nu}^{*}({\bf r}^{\prime })-v_{\nu}({\bf r})u_{\nu}^{*}
({\bf r}^{\prime }), \nonumber
\end{eqnarray}
where the index $\nu$ stands for the set of quantum numbers, and the functions
$u_{\nu}$, $v_{\nu}$ satisfy generalized Bogolyubov-De Gennes equations
\begin{eqnarray}
\varepsilon_{\nu}u_{\nu}({\bf r})\!&\!=\!&\!\left[\!-\frac{\Delta }{2}\!+\!
V({\bf r})\!-\!\mu\!+\!\!\Sigma (\!\varepsilon_{\nu}\!)\!\right]
\!\!u_{\nu}({\bf r})
\!-\!\!\Sigma _{a}(\!\varepsilon_{\nu}\!)v_{\nu}({\bf r}),
\!\!\!  \label{BDGgenu} \\
\!-\varepsilon_{\nu}v_{\nu}({\bf r})\!&\!=\!&\!\left[\!-\frac{\Delta }{2}\!
+\!V({\bf r})\!-\!\mu\!+\!\!\Sigma ^{+}\!(\!\varepsilon_{\nu}\!)\!\right]\!
\!v_{\nu}({\bf r})\!-\!
\!\Sigma_{a}(\!\varepsilon_{\nu}\!)u_{\nu}({\bf r}).\!\!\!  \label{BDGgenv}
\end{eqnarray}

In the Bogolyubov-De Gennes approach only the terms bilinear in $\widehat{%
\Psi }^{\prime }$ operators are retained in the interaction Hamiltonian 
$\hat{H}_{1}$, which assumes that the condensate density is much larger 
than the density of above-condensate particles. Then, the self-energy 
operators take the form
\begin{eqnarray}
\Sigma (\varepsilon ,{\bf r},{\bf r}^{\prime }) &=&2n_{0}\widetilde{U}\delta
({\bf r}-{\bf r}^{\prime }),  \label{sigmas0} \\
\Sigma _{a}(\varepsilon ,{\bf r},{\bf r}^{\prime }) &=&n_{0}\widetilde{U}
\delta ({\bf r}-{\bf r}^{\prime }).   \label{sigmasa0}
\end{eqnarray}
The result of their action on the condensate wavefunction $\Psi_0({\bf r})$
and the functions $u_{\nu}({\bf r})$, $v_{\nu}({\bf r})$ is reduced to
\[
\Sigma (\varepsilon_{\nu})\Psi_0({\bf r})=\int d^{3}r^{\prime}
\Sigma (\varepsilon_{\nu},{\bf r},{\bf r}^{\prime})\Psi_0({\bf r}^{\prime})
=2n_{0}({\bf r})\widetilde{U}\Psi_0({\bf r})
\]
and similar relations for the other combinations.
Then, Eq.(\ref{GPeq}) becomes the ordinary Gross-Pitaevskii equation
\begin{equation}
\left[ -\frac{\Delta }{2}+V({\bf r})-\mu +n_{0}({\bf r})\widetilde{U}\right]
\Psi _{0}({\bf r})=0,  \label{GPeq0}
\end{equation}
and Eqs.~(\ref{BCoordG}),(\ref{BCoordF}) are transformed to the ordinary
Bogolyubov-De Gennes equations
\begin{eqnarray}
\varepsilon_{\nu}u_{\nu}({\bf r}) &=&\left[ -\frac{\Delta }{2}+V({\bf r})-
\mu+2n_{0}({\bf r})\widetilde{U}\right] u_{\nu}({\bf r})-n_{0}({\bf r})
\widetilde{U}v_{\nu}({\bf r}),  \label{BDGu} \\
-\varepsilon_{\nu}v_{\nu}({\bf r}) &=&\left[ -\frac{\Delta }{2}+V({\bf r})-
\mu+2n_{0}({\bf r})\widetilde{U}\right] v_{\nu}({\bf r})-n_{0}({\bf r})
\widetilde{U}u_{\nu}({\bf r}).  \label{BDGv}
\end{eqnarray}
Taking into account Eq.(\ref{GPeq0}), in terms of the functions
$f_{\nu}^{\pm}=u_{\nu}\pm v_{\nu}$ these equations can be rewritten as
\begin{eqnarray}
\varepsilon_{\nu}f^-_{\nu}({\bf r})&=&
\left(-\frac{\Delta}{2}+\frac{\Delta\Psi_0}{2\Psi_0}\right)
f^+_{\nu}({\bf r})  \label{fplus0}\\
\varepsilon_{\nu}f^+_{\nu}({\bf r})&=&
\left(-\frac{\Delta}{2}+\frac{\Delta\Psi_0}{2\Psi_0}+2|\Psi_0|^2\widetilde{U}
\right)f^-_{\nu}({\bf r}).   \label{fmin0}
\end{eqnarray}

For a trapped Bose-condensed gas in the Thomas-Fermi regime, where $\mu
\approx n_{0\max }\widetilde{U}$ is much larger then the spacing between
the trap levels, the kinetic energy term in Eq. (\ref{GPeq0}) can be
omitted and one has \cite{algPsi}
\begin{equation}
\Psi _{0}=\sqrt{\frac{\mu -V({\bf r})}{\widetilde{U}}},  \label{psi00}
\end{equation}
if the argument of the square root is positive and zero otherwise.
For the low-energy excitations ($\varepsilon_{\nu}\ll n_{0m}\widetilde{U}$)
of Thomas-Fermi condensates Eqs.~(\ref{fplus0}),(\ref{fmin0}) can be
reduced to hydrodynamic equations obtained by Stringari \cite{Stringari}
and solved in the case of spherically symmetric harmonic potential $V(r)$
and for some excitations in a cylindrically symmetric potential.
An analytical method of solving Eqs.~(\ref{fplus0}),(\ref{fmin0}) (or the
corresponding hydrodynamic equations)
for the low-energy excitations of Thomas-Fermi condensates in an anisotropic
harmonic potential $V({\bf r})$ has been developed in \cite{Gora97,Fliesser}.

For a spatially homogeneous gas the generalized Gross-Pitaevskii equation 
(\ref{GPeq}) is equivalent to the
Pines-Hugenholtz identity \cite{PinesH}. In the Bogolyubov approach it simply
gives $\mu =n_{0}\widetilde{U}$, and Eqs.~(\ref{BDGu}),(\ref{BDGv}) lead to
the Bogolyubov spectrum
\begin{equation}
\varepsilon _{p}=\sqrt{(p^{2}/2)^{2}+n_{0}\widetilde{U}p^{2}},
\label{BogSpectrum}
\end{equation}
where ${\bf p}$ is the momentum of the excitation.

Under the condition $n_{0}\gg n^{\prime }$, for which the Bogolyubov
approach was originally developed, one can simply put $n_{0}$ equal to the
total density $n$ in Eq.(\ref{BogSpectrum}). For $n^{\prime }\gtrsim n_{0}$,
which can be the case at $T\gg \mu $, the dispersion law becomes essentially
temperature dependent \cite{LeeYang,P}. In a spatially homogeneous gas the temperature
dependence predominantly originates just from the presence of above
condensate particles, with $n^{\prime }\approx n(T/T_{c})^{3/2}$ where
$T_{c}=3.31n^{2/3}$ is the BEC transition temperature. This
leads to the replacement $n_{0}\rightarrow n_{0}+n^{\prime }$ in Eq.(\ref
{sigmas0}) and gives $\mu =(n_{0}+2n^{\prime })\widetilde{U}$. The
dispersion law will be still given by Eq.(\ref{BogSpectrum}) in which the
condensate density is now temperature dependent: $n_{0}=n\left[
1-(T/T_{c})^{3/2}\right]$.

\section{Spatially homogeneous Bose-condensed gas}

In this section we present the results for the damping rates and energy
shifts of elementary excitations in an
infinitely large spatially homogeneous Bose-condensed gas. As one can see
from Eqs.(\ref{BDGu}),(\ref{BDGv}), for finding the energy spectrum and
wavefunctions of the excitations it is sufficient to calculate the
self-energies $\Sigma $, $\Sigma ^{+}$ and $\Sigma _{a}$.
We will perform the calculations in the frequency-momentum representation
and for physical transparency consider temperatures
\begin{equation}         \label{T}
T\gg n_0\tilde U
\end{equation}
(the opposite limiting case has been discussed by Popov \cite{PopovBook} with
regard to the phonon branch of the excitation spectrum).
In the zero order approximation in the parameter
$\left( n_{0}a^{3}\right)^{1/2}$
we have the well-known mean field result: $\Sigma ^{(0)}=\Sigma
^{(0)+}=2(n_{0}+n^{\prime (0)})\widetilde{U}$, $\Sigma _{a}^{(0)}=n_{0}%
\widetilde{U}$, with $n^{\prime (0)}=n(T/T_c)^{3/2}$ (see above).
In this approach we obtain the Bogolyubov quasiparticle excitations
with the spectrum (\ref{BogSpectrum}), which we use in order to
calculate the next order in $\left( n_{0}a^{3}\right) ^{1/2}$.
The latter is determined by the contribution of diagrams containing one
quasiparticle loop \cite{PopovBook} (see Figures 1 and 2). 
Actually in this approach
we represent the Hamiltonian as the sum of the (diagonalized) Bogolyubov
Hamiltonian and the perturbation $\widehat{H}_{{\rm int}}$ originating
from $\widehat{H}_{1}$ (\ref{Hpert}) and containing the terms proportional to
$\Psi _{0}\hat\Psi ^{\prime 3}$ and $\hat\Psi^{\prime 4}$:
\begin{equation}        \label{Hint}
\hat H_{{\rm int}}=\tilde U\int d^3r[\Psi_0({\bf r})
\hat\Psi^{\prime\dagger}({\bf r})
\{\hat\Psi^{\prime\dagger}({\bf r})+\hat\Psi^{\prime}({\bf r})\}
\hat\Psi^{\prime}({\bf r})+(1/2)\hat\Psi^{\prime\dagger}({\bf r})
\hat\Psi^{\prime\dagger}({\bf r})\hat\Psi^{\prime}({\bf r})
\hat\Psi^{\prime}({\bf r})].
\end{equation}
Retaining only the temperature-dependent
contributions, after laborious calculations for the normal self-energy we
obtain $\Sigma=\Sigma^{(0)}+\Sigma^{(1)}$, where
\begin{eqnarray}
&\Sigma^{(1)}(P)&=\Sigma ^{n}(P)+\Sigma ^{r}(P),  \nonumber \\
&\Sigma^{n}(P)&=2\widetilde{U}^{2}n_{0}\int
\frac{d^{3}q}{(2\pi)^3}
(n_{q}+n_{k})\left(\frac{2A_{k}B_{q}+
A_{q}A_{k}-4A_{k}C_{q}+2C_{q}C_{k}}
{\varepsilon -\varepsilon _{q}-\varepsilon _{k}}-\right. \nonumber \\
& & \left.\frac{2A_{q}B_{k}+B_{q}B_{k}-4B_{k}C_{q}+
2C_{q}C_{k}}
{\varepsilon+\varepsilon _{k}+\varepsilon _{q}}\right)-
 8(\pi n_{0}a^{3})^{1/2}T,
\label{Sigman} \\
&\Sigma^{r}(\!P)\!\!&=2\widetilde{U}^{\!2}\!n_{\!0}\!\!\int\!\!\!
\frac{d^{3}q}{(2\pi)^3}
\!\frac{\!n_{\!q}\!-\!n_{\!k}}{\varepsilon\! +\!\varepsilon _{\!q}\!-\!
\varepsilon _{\!k}}
(2A_{q}A_{k}\!+\!2A_{k}B_{q}\!+\!2B_{k}B_{q}\!+\!
4C_{k}C_{q}\!-\!4A_{k}C_{q}\!-\!4B_{q}C_{k}).
\label{Sigmar}
\end{eqnarray}
Here $P=\{\varepsilon ,{\bf p}\}$, ${\bf k}={\bf q}+{\bf p}$,
$E_{p}=p^{2}/2$, $\varepsilon_{p}$ is given by Eq.(\ref{BogSpectrum}),
$n_{q}$ is the equilibrium occupation number,
$C_{p}=n_{0}\widetilde{U}/2\varepsilon _{p}$ and
$A_{p},B_{p}=(\pm \varepsilon _{p}+E_{p}+n_{0}\widetilde{U})/2
\varepsilon _{p}$. Similarly, the correction to the anomalous self-energy
is given by
\begin{eqnarray}
\Sigma_{a}^{(1)}(P)&=&\Sigma _{a}^{n}(P)+\Sigma _{a}^{r}(P),  \nonumber \\
\Sigma_{a}^{n}(P)&=&2\widetilde{U}^{2}n_{0}
\int\frac{d^{3}q}{(2\pi)^{3}}(n_{k}+n_{q})
\left(\frac{2A_{k}B_{q}-2A_{k}C_{q}-
2B_{q}C_{k}+3C_{q}C_{k}}
{\varepsilon -\varepsilon_{q}-\varepsilon _{k}}-\right.\nonumber\\
& & \left.
\frac{2A_{q}B_{k}-2B_{k}C_{q}-2A_{q}C_{k}+
3C_{k}C_{q}}
{\varepsilon+\varepsilon _{k}+\varepsilon _{q}}\right)-
4(\pi n_{0}a^{3})^{1/2}T,
\label{sigmaan} \\
\Sigma _{a}^{r}(P)\!&=\!\!&\!2\widetilde{U}^{2}n_{0}\int\!\!\!
\frac{d^{3}q}{(2\pi)^{3}}\frac{n_{q}-n_{k}}{\varepsilon +\varepsilon _{q}-
\varepsilon _{k}}
(2A_{\!k}A_{\!q}\!+\!2B_{\!k}B_{\!q}\!+\!6C_{\!k}C_{\!q}
\!-\!2A_{\!k}C_{\!q}\!-\!2A_{\!q}C_{\!k}\!-\!2B_{\!q}C_{\!k}
\!-\!2B_{\!k}C_{\!q}).
\label{sigmaar}
\end{eqnarray}
The resonant parts $\Sigma ^{r},\Sigma _{a}^{r}$ originate from the terms
where one of the intermediate quasiparticles is created and another one
annihilated, and the non-resonant parts $\Sigma ^{n},\Sigma _{a}^{n}$ from
the terms where both intermediate quasiparticles are created (annihilated).
Temperature independent terms in the non-resonant parts, found by Beliaev
\cite{Belyaev}, are omitted in Eqs.(\ref{Sigman})-(\ref{sigmaar}).

Each of the self-energies (\ref{Sigman})-(\ref{sigmaar}) is singular at
$P\rightarrow 0$ and at least for
small momenta the corrections become larger than the mean field values (\ref
{sigmas0}). Nevertheless, keeping in mind that any physical quantity is
determined by the combinations of the self-energies, which do not contain
the infra-red singularities, we will still treat $\Sigma ^{(1)}$ and $\Sigma
_{a}^{(1)}$ as perturbations.

For a spatially homogeneous gas the Pines-Hugenholtz identity
$\mu =(\Sigma (P)-\Sigma _{a}(P))|_{P\rightarrow 0}$
gives the first order correction to the chemical potential
\begin{equation}\label{mucorr}
\mu ^{(1)}=-\beta\sqrt{n_0};\,\,\,\,\,\,\beta=12(\pi a^3)^{1/2}T,
\end{equation}
and the relation between $n_0$ and the chemical potential, 
$\mu=n_0\widetilde{U}-\beta\sqrt{n_0}$, coincides
with that found by Popov \cite{PopovBook}. 
The $u,v$ functions in generalized Bogolyubov-De Gennes equations
(\ref{BDGgenu}), (\ref{BDGgenv}) can be written
as $u_{p}\exp (i{\bf pr})$ and $v_{p}\exp (i{\bf pr})$, and in terms
of the functions $f^{\pm}_p=u_p\pm v_p$ these equations take the form
\begin{eqnarray}
(\varepsilon-S^{-}(P))f^{-}_p&=&\left(\frac{p^2}{2}+S^{+}_{-}(P)\right)
f^{+}_p,      \label{form1}\\
(\varepsilon-S^{-}(P))f^{+}_p&=&\left(\frac{p^2}{2}+2n_0\widetilde{U}
+S^{+}_{+}(P)\right)f^{-}_p,   \label{form2}
\end{eqnarray}
where
\begin{eqnarray}
S^{+}_{\pm}&=&\frac{\Sigma^{(1)}+\Sigma^{+(1)}}{2}+\beta\sqrt{n_0}\pm
\Sigma_a,     \label{S+}\\
S^{-}&=&\frac{\Sigma^{(1)}-\Sigma^{+(1)}}{2}.  \label{S-}
\end{eqnarray}
Considering the terms $S^{-}$, $S^{+}_{\pm}$ in
Eqs.~(\ref{form1}),(\ref{form2}) as small perturbations
we put $\varepsilon=\varepsilon_p$ in the expressions for these quantities,
following from Eqs.~(\ref{Sigman})-(\ref{sigmaar}). Then,
solving Eqs.~(\ref{form1}),(\ref{form2}), for the
excitation energy we obtain $\varepsilon =\varepsilon _{p}+\varepsilon
_{p}^{(1)}$, where
\begin{equation}
\!\varepsilon _{p}^{(1)}\!\!=\!\!\left[\!\frac{E_{p}}{2\varepsilon _{p}}
\!S^{+}_{+}(P)\!+\!\frac{\varepsilon_{p}}{2E_{p}}
S^{+}_{-}(P)\!+\!S^{-}(P)\right]
_{P\rightarrow (\varepsilon _{p},{\bf p})}\!,  \label{cor}
\end{equation}

As  $\Sigma^{(1)}$ and $\Sigma _{a}^{(1)}$ are complex, the correction to
the excitation energy has both a real and an imaginary part: $\varepsilon
_{p}^{(1)}=\delta \varepsilon _{p}-i\Gamma _{p}$.
The former gives the energy shift, and the latter is responsible for
damping of the excitations.
Under the condition $T\gg n_{0}\widetilde{U}$ a straightforward calculation
of Eq.(\ref{cor}) on the basis of Eqs.~(\ref{Sigman})-(\ref{sigmaar}) and
(\ref{S+}),(\ref{S-}), for the phonon branch of the spectrum
($\varepsilon _{p}\ll n_{0}\widetilde{U}$) yields
\begin{eqnarray}
\delta \varepsilon _{p} &\approx &-7\varepsilon _{p}\frac{T}
{n_{0}\widetilde U}(n_{0}a^{3})^{1/2},  \label{reshomRe} \\
\Gamma _{p} &=&\varepsilon _{p}\frac{3\pi ^{3/2}T}
{4n_{0}\widetilde U}(n_{0}a^{3})^{1/2}.  \label{reshomIm}
\end{eqnarray}
It is important to emphasize that in this case the energy shift is
determined by both resonant and non-resonant terms, whereas the damping rate
$\Gamma _{p}$ is
described solely by the resonant contributions. This type of damping
originates from quasi-resonant scattering of thermal excitations from a given
excitation (Landau damping) and is absent at $T=0$. Both the energy shift
and the damping rate are determined by the interaction of a given excitation
with intermediate quasiparticles having energies
$\varepsilon _{q}\sim n_{0}\widetilde{U}$.
The damping rate $\Gamma _{p}$ (\ref{reshomIm}) coincides
with that found in recent contributions \cite{HS,Recent,FSW,Recent1} and contains 
a slight numerical difference from the earlier Szepfalusy-Kondor result \cite{SK}. The energy shift for the phonon branch of the spectrum was also calculated
in \cite{HS}. In the latter work the expansion of the self-energy functions
near the point $\varepsilon=\varepsilon_{p}$ was used and formally divergent
integrals were canceling each other in the final expression for the energy 
shift, which have led to the result by approximately factor 6 smaller then 
the shift (\ref{reshomRe}) obtained by the exact integration.

Eqs.(\ref{reshomRe}),(\ref{reshomIm}) clearly show that the small
parameter of the theory is
$(T/n_0\tilde U)(n_{0}a^{3})^{1/2}\ll 1$ (see Eq.(\ref{small})),
whereas in the zero temperature approach \cite{Belyaev} the small parameter is
$(n_{0}a^{3})^{1/2}\ll 1$.
The presence of an additional large factor $T/n_0\widetilde{U}$ at finite
temperatures $T\gg n_0\widetilde{U}$ originates from the Bose enhancement 
diagrams containing one quasiparticle loop: Compared to the zero-temperature 
case the contribution of each of these diagrams is multiplied by the Bose
factor $n_{q}=[\exp{(\varepsilon_q/T)}-1]^{-1}$ (or $1+n_{q}$). As the most 
important is the contribution of intermediate quasiparticles with energies 
$\varepsilon_q\sim n_0\widetilde{U}$, for $T\gg n_0\widetilde{U}$ the Bose
factor $n_q\sim T/n_0\widetilde{U}$.
The criterion similar to Eq.(\ref{small}) was found by Popov \cite{P,PopovBook}
as the condition which allows one to use the mean field approach at finite 
temperatures and to renormalize the theory for reaching beyond this approach.

Remarkably, the criterion (\ref{small}) is fulfilled even at temperatures
very close to $T_{c}$. For $\Delta T=T_{c}-T\ll T_{c}$ we have $n_{0}\sim
n\Delta T/T_{c}$, and Eq.(\ref{small}) gives $\Delta T/T_{c}\gg
(n_{0}a^{3})^{1/3}$, which coincides with the well known Ginzburg criterion
\cite{Ginzburg} for the absence of critical fluctuations.
The criterion (\ref{small}) also ensures that the main contribution to the
damping rate originates from the interaction of a given excitation with
thermal excitations through the condensate, i.e., from the first term
in $\hat H_{\rm int}$ (\ref{Hint}).

At energies $\varepsilon_p\agt n_0\widetilde U$ (but still
$\varepsilon_p\ll T$) the results for the energy shift and damping rate,
following from Eq.(\ref{cor}), can be represented in the form
\begin{eqnarray}
\delta\varepsilon_p&=&\varepsilon_pF_1\left(\frac{\varepsilon_p}
{n_0\widetilde U}\right)\frac{T}{n_0\widetilde U}(n_0a^3)^{1/2},
\label{homRe} \\
\Gamma_p&=&\varepsilon_pF_2\left(\frac{\varepsilon_p}
{n_0\widetilde U}\right)\frac{T}{n_0\widetilde U}(n_0a^3)^{1/2}.
\label{homIm}
\end{eqnarray}
The functions $F_1$ and $F_2$ have been calculated numerically. For
$\varepsilon_p\ll n_0\widetilde U$ we have $F_1\approx -7$,
$F_2=3\pi^{3/2}/4$ and Eqs.~(\ref{homRe}), (\ref{homIm}) coincide
with Eqs.~(\ref{reshomRe}), (\ref{reshomIm}). The dependence of
$\delta\varepsilon_p$ and $\Gamma_p$ on the excitation energy is presented
in Fig.3.
For $\varepsilon_p\alt n_0\widetilde U$ the damping rate increases with
$\varepsilon_p$ and reaches its maximum at
$\varepsilon_p\sim 10n_0\widetilde U$. Further increase of $\varepsilon_p$
leads to slowly decreasing $\Gamma_p$. For single-particle excitations
($\varepsilon_p=p^2/2\gg n_0\widetilde U$) we obtain
$F_2(\varepsilon_p/n_0\widetilde U)\sim (n_0\widetilde U/\varepsilon_p)^{3/2}$.
Accordingly, the damping rate can be written as $\Gamma_p\sim (T/\varepsilon_p)
n_0\sigma v_p$, where $\sigma=8\pi a^2$ is the elastic cross section,
and $v_p$ the particle velocity.
This damping rate exceeds the Beliaev temperature-independent term
$n\sigma v_p$ even at $T$ close to $T_c$, if $\Delta T=T_c-T\gg \varepsilon_p$.
In contrast to the phonon branch of the
spectrum, for $\varepsilon_p\agt n_0\widetilde U$ the damping is provided
by both the Szepfalusy-Kondor ($\nu+\gamma\leftrightarrow\gamma'$) and
Beliaev ($\nu\leftrightarrow\gamma+\gamma'$) processes and, hence, can
no longer be treated as Landau damping. The small parameter of the
theory is still given by Eq.(\ref{small}), since even at
$\varepsilon_p\gg n_0\widetilde U$ the energy of at least one of the thermal
excitations is of order $n_0\widetilde U$.

\begin{figure}
\epsfig{file=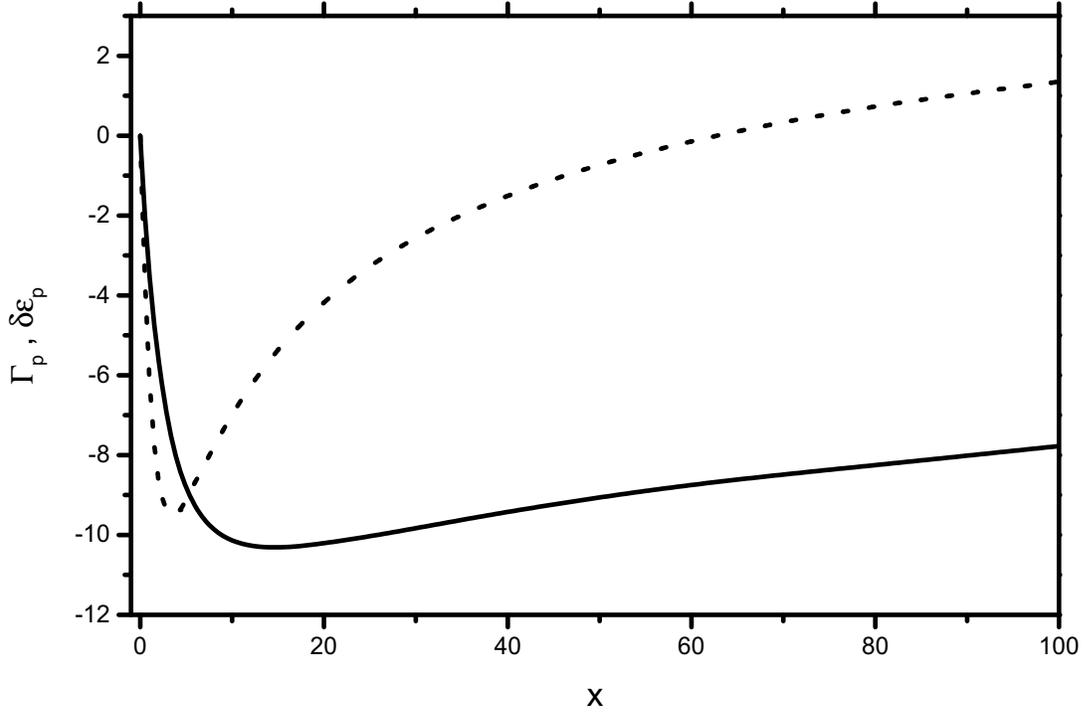, width=0.9\linewidth}
\caption{ The damping rate $\Gamma_p$ (solid line) and the energy shift
$\delta\varepsilon_p$ (dashed line) versus $(\varepsilon_p/n_0
\widetilde U)$. Both $\Gamma_p$ and $\delta\varepsilon_p$ are given in the
units of $T(n_0a^3)^{1/2}$.
}
\label{f3}
\end{figure}

The energy shift for $\varepsilon_p\alt n_0\widetilde{U}$ is negative.
The modulus of the shift increases with $\varepsilon_p$ and reaches
its maximum at $\varepsilon_p\approx 4n_0\widetilde{U}$. The further
increase of $\varepsilon_p$ decreases $|\delta\varepsilon_p|$. The
latter is equal to zero for $\varepsilon_p\approx 60n_0\widetilde{U}$,
and becomes positive at larger $\varepsilon_p$.

The above results for the damping rate and energy shift of a given excitation 
are obtained in the so called collisionless regime: We assume that the De Broglie
wavelength of the excitation, $1/p$, is much larger than the mean free path of the
thermal quasiparticles with energies $\sim n_0\widetilde{U}$, which are mostly 
responsible for the damping and shifts.
It is also assumed that the excitation energy $\varepsilon_p$ greatly exceeds the
damping rate of these thermal excitations. The latter is of order $T(n_0a^3)^{1/2}$
(see Fig.3), and for $\varepsilon_p\agt n_0\widetilde{U}$ the two requirements of
the collisionless regime are well satisfied under condition (\ref{small}).
In the phonon branch of the excitation spectrum ($\varepsilon_p\ll n_0\widetilde{U}$)
these requirements are equivalent to each other, and the collisionless criterion
can be simply written as
\begin{equation}      \label{cols}
\varepsilon_p\gg T(n_0a^3)^{1/2}.
\end{equation}
As clearly seen, in the phonon branch one can always find excitations
which do not satisfy Eq.(\ref{cols}) and, hence, require a hydrodynamic
description with regard to their damping rates and energy shifts.

The collisionless criterion (\ref{cols}) provides an additional argument
on support of the above used perturbative approach for solving
Eqs.~(\ref{form1}), (\ref{form2}). Under condition (\ref{cols}) the
term $S^{-}\alt T(n_0a^3)^{1/2}\ll\varepsilon_p$, the
term $S^{+}_{-}\alt\varepsilon_p(T/n_0\widetilde{U})(n_0a^3)^{1/2}\ll p^2$,
and the term $S^{+}_{+}\alt (n_0\widetilde{U}/\varepsilon_p)
T(n_0a^3)^{1/2}\ll n_0\widetilde{U}$.

The non-mean-field shift $\delta\varepsilon_p$ is actually the shift of
the excitation energy $\varepsilon_{p}$ at a given condensate density $n_0$.
On the other hand, $\varepsilon_p$ is determined by the Bogolyubov
dispersion law (\ref{BogSpectrum}),
with the temperature-dependent condensate density $n_0(T)$, and, hence,
is temperature-dependent by itself.
Therefore, at a given $T$ one will also have the mean-field
temperature-dependent energy shift
$\delta\varepsilon_p^{\rm mf}=\varepsilon_p(T)-\varepsilon_p(0)$.
As the condensate density decreases with increasing temperature,
$\delta\varepsilon_p^{\rm mf}$ is always negative.
For $T\gg n_0\widetilde{U}$ it greatly exceeds the above calculated
shift $\delta\varepsilon_p$ at any $p$.
The ratio $(\delta\varepsilon_p^{\rm mf}/\delta\varepsilon_p)$ decreases
with temperature, but even for $n'\ll n_0$ one has
\begin{equation}       \label{mfs}
\delta\varepsilon_p^{\rm mf}=-\varepsilon_p(0)\frac{n'(T)\widetilde{U}}
{p^2/2+2n_0(0)\widetilde{U}},
\end{equation}
and $(\delta\varepsilon_p^{\rm mf}/\delta\varepsilon_p)\sim (T/n_0
\widetilde{U})^{1/2}\gg 1$.

\section{Spatially inhomogeneous Bose-condensed gas}

We now generalize the above obtained results for the case of
elementary excitations in a spatially inhomogeneous (trapped)
Bose-condensed gas.
As already mentioned in the introduction, a new ingredient here is related
to the inhomogeneous density profile of the condensate and the discrete
structure of the excitation spectrum.
This requires us to develop a new method of calculating the self-energy
functions in generalized Bogolyubov-De Gennes equations (\ref{BDGgenu}),
(\ref{BDGgenv}).
The self-energy operators in these equations are the sums of the
zeroth and first order terms:
\begin{eqnarray}
\Sigma_a(\varepsilon,{\bf r},{\bf r'})&
=&n_0({\bf r})\widetilde{U}\delta({\bf r}-{\bf r'})+\Sigma_a^{(1)},
\label{0}\\
\Sigma(\varepsilon,{\bf r},{\bf r'})&=&2(n_0({\bf r})+n^{\prime(0)})
\widetilde{U}\delta({\bf r}-{\bf r'})+\Sigma^{(1)},
\label{00}
\end{eqnarray}
and a similar
relation for $\Sigma^{+}$. At temperatures $T\gg n_{0m}\widetilde{U}$
the zero order value of the above-condensate density in the condensate
spatial region is coordinate independent and equal to the above-condensate 
density in the ideal gas approach: $n^{\prime(0)}(T)=2.6(T/2\pi)^{3/2}$. 
On the contrary, the self-energies $\Sigma^{(1)}$,
$\Sigma_a^{(1)}$ depend explicitly on the condensate density.
Due to the discrete structure of the energy spectrum of excitations the
expressions for these quantities should be written in the form of sums
over the discrete states of intermediate quasiparticles $\gamma$,$\gamma'$.
In the frequency-coordinate representation we have
\begin{eqnarray}
& &\Sigma(\varepsilon_{\nu},{\bf r},{\bf r}')=\Sigma^n
(\varepsilon_{\nu},{\bf r},{\bf r}')+\Sigma^r
(\varepsilon_{\nu},{\bf r},{\bf r}'),  \nonumber\\
& &  \nonumber\\
& &\Sigma^n(\varepsilon_{\nu},{\bf r},{\bf r}')=2n_0\tilde U^2
\sum_{\gamma,\gamma'}(n_{\gamma}+n_{\gamma'})
\left(\frac{2u_{\gamma}({\bf r})u_{\gamma}({\bf r}')
v_{\gamma'}({\bf r})v_{\gamma'}({\bf r}')+u_{\gamma}({\bf r})u_{\gamma}({\bf r}')
u_{\gamma'}({\bf r})u_{\gamma'}({\bf r}')
}{\varepsilon-\varepsilon_{\gamma}-\varepsilon_{\gamma'}}+\right.\nonumber\\
& &\frac{
-4u_{\gamma}({\bf r})u_{\gamma}({\bf r}')u_{\gamma'}({\bf r})v_{\gamma'}({\bf r})+2u_{\gamma}({\bf r})v_{\gamma}({\bf r}')
u_{\gamma'}({\bf r})v_{\gamma'}({\bf r}')}
{\varepsilon-\varepsilon_{\gamma}-\varepsilon_{\gamma'}}-
\nonumber\\
& &\frac{2v_{\gamma}({\bf r})v_{\gamma}({\bf r}')u_{\gamma'}({\bf r})u_{\gamma'}({\bf r}')+
v_{\gamma}({\bf r})v_{\gamma}({\bf r}')v_{\gamma'}({\bf r})v_{\gamma'}({\bf r}')-
4v_{\gamma}({\bf r})v_{\gamma}({\bf r}')u_{\gamma'}({\bf r})v_{\gamma'}({\bf r}')}{\varepsilon+\varepsilon_{\gamma}+\varepsilon_{\gamma'}}\nonumber \\+
& &\left.\frac{2u_{\gamma}({\bf r})v_{\gamma}({\bf r}')u_{\gamma'}({\bf r})v_{\gamma'}({\bf r}')}
{\varepsilon+\varepsilon_{\gamma}+\varepsilon_{\gamma'}}\right)
-8(n_0\pi a^3)^{1/2}T\delta({\bf r}-{\bf r}'),\label{TNSn}\\
& &  \nonumber\\
& &\Sigma^r(\varepsilon_{\nu},{\bf r},{\bf r}')=2n_0\tilde U^2
\sum_{\gamma,\gamma'}\frac{n_{\gamma'}-n_{\gamma}}{\varepsilon+
\varepsilon_{\gamma'}-\varepsilon_{\gamma}}\nonumber\\
& &(2u_{\gamma}({\bf r})u_{\gamma}({\bf r}')
u_{\gamma'}({\bf r})u_{\gamma'}({\bf r}')+2u_{\gamma}({\bf r})u_{\gamma}({\bf r}')
v_{\gamma'}({\bf r})v_{\gamma'}({\bf r}')
+2v_{\gamma}({\bf r})v_{\gamma}({\bf r}')v_{\gamma'}({\bf r})v_{\gamma'}({\bf r}')+\nonumber\\
& &4u_{\gamma}({\bf r})v_{\gamma}({\bf r}')u_{\gamma'}({\bf r})v_{\gamma'}({\bf r}')-
4u_{\gamma}({\bf r})u_{\gamma}({\bf r}')u_{\gamma'}({\bf r})v_{\gamma'}({\bf r}')-
4u_{\gamma}({\bf r})v_{\gamma}({\bf r}')v_{\gamma'}({\bf r})
v_{\gamma'}({\bf r}')).\label{TRSn}\\
& &  \nonumber\\
& &\Sigma_a=\Sigma_a^n(\varepsilon_{\nu},{\bf r},{\bf r}')+
\Sigma_a^r(\varepsilon_{\nu},{\bf r},{\bf r}'), \nonumber\\
& &  \nonumber\\
& &\Sigma_a^n(\varepsilon_{\nu},{\bf r},{\bf r}')=2n_0\tilde U^2
\sum_{\gamma,\gamma'}(n_{\gamma}+n_{\gamma'})\nonumber\\
& &\left(\frac{2u_{\gamma}({\bf r})u_{\gamma}({\bf r}')
v_{\gamma'}({\bf r})v_{\gamma'}({\bf r}')-2u_{\gamma}({\bf r})u_{\gamma}({\bf r}')
u_{\gamma'}({\bf r})v_{\gamma'}({\bf r}')}
{\varepsilon-\varepsilon_{\gamma}-\varepsilon_{\gamma'}}\right.\nonumber\\
& &\frac{-2u_{\gamma}({\bf r})v_{\gamma}({\bf r}')v_{\gamma'}({\bf r})v_{\gamma'}({\bf r}')+
3u_{\gamma}({\bf r})v_{\gamma}({\bf r}')
u_{\gamma'}({\bf r})v_{\gamma'}({\bf r}')}
{\varepsilon-\varepsilon_{\gamma}-\varepsilon_{\gamma'}}-
\nonumber\\
& &\frac{2v_{\gamma}({\bf r})v_{\gamma}({\bf r}')u_{\gamma'}({\bf r})u_{\gamma'}({\bf r}')-
2v_{\gamma}({\bf r})v_{\gamma}({\bf r}')u_{\gamma'}({\bf r})v_{\gamma'}({\bf r}')}{\varepsilon+\varepsilon_{\gamma}+\varepsilon_{\gamma'}}\nonumber\\
& &-\left.\frac{
2u_{\gamma}({\bf r})v_{\gamma}({\bf r}')u_{\gamma'}({\bf r})u_{\gamma'}({\bf r}')+
3u_{\gamma}({\bf r})v_{\gamma}({\bf r}')u_{\gamma'}({\bf r})v_{\gamma'}({\bf r}')}
{\varepsilon+\varepsilon_{\gamma}+\varepsilon_{\gamma'}}\right)
-4(n_0\pi a^3)^{1/2}T\delta({\bf r}-{\bf r}'),\label{TNSa}\\
& &  \nonumber\\
& &\Sigma_a^r(\varepsilon_{\nu},{\bf r},{\bf r}')=2n_0\tilde U^2
\sum_{\gamma,\gamma'}\frac{n_{\gamma'}-n_{\gamma}}{\varepsilon+
\varepsilon_{\gamma'}-\varepsilon_{\gamma}}\nonumber\\
& &(2u_{\gamma}({\bf r})u_{\gamma}({\bf r}')
u_{\gamma'}({\bf r})u_{\gamma'}({\bf r}')+2v_{\gamma}({\bf r})v_{\gamma}({\bf r}')
v_{\gamma'}({\bf r})v_{\gamma'}({\bf r}')
+6u_{\gamma}({\bf r})v_{\gamma}({\bf r}')u_{\gamma'}({\bf r})v_{\gamma'}({\bf r}')-\nonumber\\
& &2u_{\gamma}({\bf r})u_{\gamma}({\bf r}')u_{\gamma'}({\bf r})v_{\gamma'}({\bf r}')-
2u_{\gamma}({\bf r})v_{\gamma}({\bf r}')u_{\gamma'}({\bf r})u_{\gamma'}({\bf r}')-
2u_{\gamma}({\bf r})v_{\gamma}({\bf r}')v_{\gamma'}({\bf r})v_{\gamma'}({\bf r}')-\nonumber\\
& &2v_{\gamma}({\bf r})v_{\gamma}({\bf r}')u_{\gamma'}({\bf r})
v_{\gamma'}({\bf r}')).\label{TRSa}
\end{eqnarray}

As we saw in the previous section, in the spatially homogeneous case
all physical quantities are
determined by the contribution to the self-energy functions
$\Sigma^{(1)}$, $\Sigma_a^{(1)}$ from
intermediate quasiparticles with energies of order the mean field
interaction between particles.
The same holds for a spatially inhomogeneous (trapped) Bose-condensed
gas in the Thomas-Fermi regime, where the mean field interaction
$n_{0m}\widetilde{U}$ greatly
exceeds the level spacing in the trapping potential.
The intermediate quasiparticles with energies of order
$n_{0m}\widetilde{U}$ are essentially quasiclassical.
With regard to the integral operator $(\Sigma^{(1)}-
\Sigma_a^{(1)})_{\varepsilon\rightarrow 0}$
in the generalized Gross-Pitaevskii equation (\ref{GPeq}), which is solely
determined by non-resonant contributions, this immediately allows
one to replace the summation over the discrete intermediate states
by integration.
The kernel of this integral operator varies
at distances $|{\bf r}-{\bf r'}|$ of order the correlation length
$l_{{\rm cor}}=1/\sqrt{n_0\widetilde{U}}$ which is much smaller than the
characteristic size of the condensate. Therefore, the result of the
operator action on the condensate wavefunction can be written in the
local density approximation and, hence, should rely on Eq.(\ref{mucorr})
with coordinate-dependent condensate density $n_0({\bf r})$:
\begin{equation}        \label{missed}
\left[\Sigma^{(1)}-\Sigma_a^{(1)}\right]_{\varepsilon\rightarrow 0}
\Psi_0({\bf r})=
\int d^3r'\left[\Sigma^{(1)}(\varepsilon,{\bf r},{\bf r'})-
\Sigma_a^{(1)}(\varepsilon,{\bf r},{\bf r'})\right]_{\varepsilon
\rightarrow 0}\Psi_0({\bf r'})=-\beta\Psi_0^2({\bf r}).
\end{equation}
This result can be easily obtained from Eqs.~(\ref{TNSn}), (\ref{TRSa}),
where one should put $\varepsilon_{\nu}=0$, neglect the difference
between $\varepsilon_{\gamma}$ and $\varepsilon_{\gamma'}$, and
make a summation over $\gamma'$. Replacing the summation over $\gamma$
by integration one should also take into account that for quasiclassical
excitations the functions $f^{\pm}_{\gamma}$ can be represented in the form
\begin{equation}        \label{form}
f^{\pm (0)}_{\gamma}({\bf r})\!=\!\left(\!\frac{\sqrt
{\!\varepsilon_{\gamma}^2\!+\!(n_0({\bf r})\tilde U)^2}
\!-\!n_0({\bf r})\tilde U}
{\varepsilon_{\gamma}}\right)^{\!\mp 1/2}\!\!\!\!f_{\gamma}({\bf r}),
\end{equation}
where $|f_{\gamma}({\bf r})|^2$ is the ratio of the local to total density
of states for Bogolyubov quasiparticles of a given symmetry, described by
the classical Hamiltonian
\begin{equation}      \label{Hcl}
H({\bf p},{\bf r})=\sqrt{(p^2/2)^2+\tilde n_0({\bf r})\tilde U p^2}.
\end{equation}

On the basis of Eq.(\ref{missed}) we obtain the generalized
Gross-Pitaevskii equation in the form
\begin{equation}          \label{GPeq1}
\left( -\frac{\Delta}{2}+V({\bf r})-\tilde \mu+\widetilde{U}|\Psi_0|^2-
\beta\Psi_0\right) \Psi_0=0,
\end{equation}
where $\tilde\mu=\mu-2n^{\prime(0)}(T)\tilde U$ is coordinate independent.
Compared to the ordinary Gross-Pitaevskii equation (\ref{GPeq0}), 
Eq.(\ref{GPeq1}) contains an extra term $[2n^{\prime (0)}\widetilde{U}-
\beta\Psi_0]\Psi_0$ in the lhs. 
One can easily check that Eq.(\ref{GPeq1}) coincides with the equation
$$
\left[-\frac{\Delta}{2}+V({\bf r})-\mu+\widetilde{U}|\Psi_0|^2+
\widetilde{U}(2\langle\hat\Psi^{\prime\dagger}\hat\Psi^{\prime}
\rangle +\langle\Psi^{\prime}\Psi^{\prime}\rangle)\right]\Psi_0=0
$$
obtained by averaging the non-linear Schr\"odinger equation for the field
operator. As mentioned above, for $T\gg n_0\widetilde{U}$ the above-condensate 
density $n^{\prime}=\langle\Psi^{\prime\dagger}\Psi^{\prime}\rangle$ in the 
condensate spatial region is mainly determined by the coordinate-independent 
(ideal gas) contribution $n^{\prime (0)}(T)$.
The coordinate-dependent correction to the above-condensate density,
$n^{\prime(1)}({\bf r})=\langle\hat\Psi^{\prime\dagger}({\bf r})
\hat\Psi^{\prime}({\bf r})\rangle -n^{\prime(0)}$, turns out to be equal to the 
anomalous average $\langle\Psi({\bf r})\Psi({\bf r})\rangle$:
$$
n^{\prime(1)}({\bf r})=\langle\Psi({\bf r})\Psi({\bf r})\rangle=
-\frac{\beta\sqrt{n_0({\bf r})}}{3\widetilde{U}}.
$$ 
Accordingly, the quantity $[2n^{\prime (0)}\widetilde{U}-\beta\Psi_0]\Psi_0=
\widetilde{U}(2\langle\hat\Psi^{\prime\dagger}\hat\Psi^{\prime}
\rangle +\langle\Psi^{\prime}\Psi^{\prime}\rangle)$.

Taking advantage of Eqs~(\ref{0}), (\ref{00}) and (\ref{GPeq1}), the
generalized Bogolyubov-De Gennes equations (\ref{BDGgenu}),
(\ref{BDGgenv}) are reduced to
\begin{eqnarray}
(\varepsilon_{\nu}-S^{-}) f^-_{\nu}({\bf r})&=&
\left(-\frac{\Delta}{2}+\frac{\Delta\Psi_0}{2\Psi_0}+S^+_-
\right)f^+_{\nu}({\bf r})  \label{fplus}\\
(\varepsilon_{\nu}-S^-) f^+_{\nu}({\bf r})&=&
\left(-\frac{\Delta}{2}+\frac{\Delta\Psi_0}{2\Psi_0}+2|\Psi_0|^2\widetilde{U}
+S^+_+ \right)f^-_{\nu}({\bf r})   \label{fmin},
\end{eqnarray}
where the quantities $S^+_{\pm}$, $S^-$ are given by Eqs.~(\ref{S+}),
(\ref{S-}), and $\varepsilon_{\nu}$ is the exact value of the excitation
energy.
Eqs.~ (\ref{GPeq1}), (\ref{fplus}) and (\ref{fmin}) represent a complete
set of equations for finding the energy shifts and damping rates of the
elementary excitations.

A precise calculation of the self-energy functions in
Eqs.~(\ref{fplus}), (\ref{fmin}) depends on the value of
$\varepsilon_{\nu}$ and on the trapping geometry.
In this section we will make general statements on how the calculation
can be performed.
In most of the cases (except the case of the lowest excitations with
zero orbital angular momentum in spherically symmetric traps) the
characteristic time scale in the self-energy operators,
$1/\varepsilon_{\nu}$, is much smaller than the inverse level spacing
in the trap. Therefore, the summation over the discrete intermediate
states can be replaced by integration.
This is a direct consequence of the general statement that the
time-dependent discrete Fourier sum can be replaced by its integral
representation at times much smaller than the inverse frequency spacing
(see e.g., \cite{Chirikov}).

The kernels of the non-resonant parts of the
self-energy operators, (\ref{TNSn}) and (\ref{TNSa}), vary
at distances $|{\bf r}-{\bf r'}|$ which do not exceed the correlation
length $l_{{\rm cor}}=1/\sqrt{n_0\widetilde{U}}$. As $l_{\rm cor}$ is 
much smaller than the characteristic size of the condensate, the 
non-resonant parts of the self-energies can be calculated in the local 
density approximation.
One can easily find from Eqs.~(\ref{TNSn}), (\ref{TNSa}) that they
lead to the result of Eqs.~(\ref{Sigman}), (\ref{sigmaan}) (with $n_0$
replaced by the coordinate-dependent density $n_0({\bf r})$),
multiplied by $\delta({\bf r}-{\bf r'})$.
For the lowest excitations $\nu$ one should put ${\bf k}={\bf q}$
in Eqs.~(\ref{Sigman}),(\ref{sigmaan}), and for quasiclassical
excitations $\nu$
take $p$ from the Bogolyubov dispersion law
$H({\bf p},{\bf r})=\varepsilon_{\nu}$.

The calculation of the resonant contributions to the self-energies
is more subtle. Using Eq.(\ref{form}) for the functions $u_{\nu},v_{\nu}=
f_{\nu}^{\pm}$ in Eqs.~(\ref{TRSn}), (\ref{TRSa}), one can see that
all resonant contributions contain the quantity
$$
Q({\bf r},{\bf r}')=\sum_{\gamma'}\frac{f_{\gamma}({\bf r})
f_{\gamma}({\bf r}')f_{\gamma'}({\bf r})f_{\gamma'}({\bf r}')}
{\varepsilon_{\nu}+\varepsilon_{\gamma}-\varepsilon_{\gamma'}+i0}.
$$
Writing $(\varepsilon_{\nu}+\varepsilon_{\gamma}-
\varepsilon_{\gamma'}+i0)^{-1}$ as the integral over time
$\int_0^{\infty}dt\exp\{i(\varepsilon_{\nu}+\varepsilon_{\gamma}-
\varepsilon_{\gamma'}+i0)t\}$, we obtain
\begin{equation}     \label{Q}
Q({\bf r},{\bf r}')=
i\int_0^{\infty}dt\exp{(i\varepsilon_{\nu}t)}K_{\gamma}({\bf r},
{\bf r}',t),
\end{equation}
where  the quantum-mechanical correlation function
\begin{equation}      \label{K1}
K_{\gamma}({\bf r},{\bf r}',t)=\sum_{\gamma'}f_{\gamma}({\bf r})
f_{\gamma}({\bf r}')f_{\gamma'}({\bf r})f_{\gamma'}({\bf r}')
\exp\left\{i(\varepsilon_{\gamma}-\varepsilon_{\gamma'}+i0)t\right\}.
\end{equation}
We will turn from the integration over the quantum states $\gamma'$
of the quasiclassical thermal
excitations to the integration along the classical trajectories of motion of
Bogolyubov-type quasiparticles in the trap. Following a general method (see
\cite{Shapoval,Gorkov,Book}), employed in \cite{FSW} for the damping
of low-energy excitations, we obtain
\begin{equation}   \label{K2}
K_{\gamma}({\bf r},{\bf r}',t)=g_{\gamma}^{-1}
\int \delta({\bf r}''-{\bf r})
\delta({\bf r}''_p(t)-{\bf r}')
\delta(\varepsilon_{\gamma}-H({\bf p},{\bf r}''))
\frac{d^3r''d^3p}{8\pi^3},
\end{equation}
where ${\bf r}_p(t)$ is the coordinate of the classical
trajectory with initial momentum ${\bf p}$ and coordinate ${\bf r}$.
Eq.(\ref{K2}) will be used in the next sections where we demonstrate the
facilities of the theory.

Concluding this section, we emphasize the key role of harmonicity of the
trapping potential for temperature-dependent energy shifts of the
excitations.
As mentioned in the previous section, in the spatially homogeneous case
at a given temperature the non-mean-field shift is much smaller than the
shift $\delta\varepsilon^{\rm mf}$ appearing in the mean field approach
simply due to the temperature dependence of the condensate density in the
Bogolyubov dispersion law (\ref{BogSpectrum}).
For the Thomas-Fermi condensate in a harmonic confining
potential the situation is different. In this case the spectrum of
low-energy ($\varepsilon_{\nu}\ll n_{0m}\widetilde{U}$) excitations
is independent of the mean field interparticle interaction
$n_{0m}\widetilde{U}$ (chemical potential) and the condensate density
profile\cite{Stringari,Gora97,Fliesser}. Hence, the
temperature-dependent energy shifts can only appear due to
non-Thomas-Fermi corrections.
For finding these corrections one should use the mean field self-energies
$\Sigma_a(\varepsilon,{\bf r},{\bf r'})=n_0({\bf r})\widetilde{U}
\delta({\bf r}-{\bf r'})$, $\Sigma(\varepsilon,{\bf r},{\bf r'})=
2(n_0({\bf r})+n^{\prime(0)})\widetilde{U}\delta({\bf r}-{\bf r'})$, where
the only difference from the $T=0$ case is related to the presence
of above-condensate particles in the condensate spatial region at finite $T$ 
through the coordinate-independent term
$2n^{\prime(0)}\widetilde{U}$ in $\Sigma$.
Then Eqs.~(\ref{fplus}),
(\ref{fmin}) take the form of ordinary Bogolyubov-De Gennes equations
(\ref{fplus0}),(\ref{fmin0}), and Eq.(\ref{GPeq1}) becomes the ordinary
Gross-Pitaevskii equation (\ref{GPeq0}), with the chemical potential $\mu$
replaced by $\tilde\mu$. The latter circumstance changes the condensate
wavefunction compared to that at $T=0$ and ensures the temperature
dependence of $\Psi_0$. Accordingly, the excitation energies
$\varepsilon_{\nu}$ in Eqs.~(\ref{fplus0}),(\ref{fmin0}) also become
temperature dependent.
This type of approach, which for a spatially homogeneous gas would
immediately lead to the result of Lee and Yang \cite{LeeYang}, has been
used in recent numerical calculations
of the energy shifts of the lowest quadrupole
excitations in spherically symmetric \cite{Hut} and cylindrically symmetric 
\cite{Dodd,Zheng} harmonic traps.
The presence of the coordinate-dependent part of the above-condensate density,
$n^{\prime (1)}({\bf r})$, in these calculations is not adequate, since the
anomalous average equal to this part was omitted and equations for the excitations 
did not contain the corrections to the self-energies, also proportional to 
$(n_0a^3)^{1/2}$. 
However, at $T\gg n_0\widetilde{U}$, where $n^{\prime (1)}\ll n^{\prime (0)}$,
the coordinate-dependent part $n^{\prime (1)}({\bf r})$ as itself should not
significantly influence the result, and the calculations \cite{Hut,Dodd,Zheng}
should actually demonstrate how important are the mean field non-Thomas-Fermi 
effects.
The results of \cite{Dodd,Zheng} show the absence of energy shifts of the 
excitations at temperatures $T<0.6T_c$ in the JILA experiment \cite{JILAshifts} 
and in this sense agree with the experimental data, but do not describe the 
upward and downward shifts of the excitation energies, observed experimentally 
at higher temperatures (in this respect it is worth mentioning that the calculations 
\cite{Shenoy} performed for the thermal cloud in the hydrodynamic regime agree
surprisingly well with the experiment \cite{JILAshifts}).
On the other hand, the calculation \cite{Zheng} shows a downward shift of the energy
of the lowest quadrupole excitation with increasing temperature in the conditions 
of the MIT experiment \cite{MITshifts}. This is consistent with the experimental data 
and indicates that for not very small Thomas-Fermi parameter 
$\omega/n_{0m}\widetilde{U}$ the mean field non-Thomas-Fermi effects can be important 
for temperature-dependent shifts of the lowest excitations.  

Below we will assume a sufficiently small Thomas-Fermi parameter
$\omega/n_{0m}\widetilde{U}$ and demonstrate the use of the
theory by the examples where the influence of non-Thomas-Fermi effects
on the energy shifts of the excitations is not important.

\section{Quasiclassical excitations in a trapped Bose-condensed gas}

We will discuss the Thomas-Fermi condensates in a harmonic confining
potential on the basis of Eqs.~(\ref{GPeq1}), (\ref{fplus}), (\ref{fmin}).
Neglecting the kinetic energy term in Eq.(\ref{GPeq1}), we arrive at a
quadratic equation for $\Psi_0$. Expanding the solution of this equation in
powers of $\beta$ and retaining only the terms independent of $\beta$ and
the terms linear in $\beta$, for the condensate density we obtain
\begin{equation}           \label{cd1}
n_0({\bf r})=\tilde n_0({\bf r})+\beta\sqrt{\tilde n_0({\bf r})}/
\widetilde{U},
\end{equation}
where $\tilde n_0({\bf r})=(\tilde \mu-V({\bf r}))/
\widetilde U$ is the density of the Thomas-Fermi condensate in the ordinary
mean field approach.

We first consider the damping and energy shifts of quasiclassical
($\varepsilon_{\nu}\gg\omega$)
low-energy excitations of a trapped Thomas-Fermi condensate, i.e., the
quasiclassical excitations with energies much smaller than the mean field
interaction between particles $n_{0m}\widetilde{U}$.
In this case the terms in Eqs.~(\ref{fplus}) and
(\ref{fmin}), originating from the kinetic energy of the condensate, can
be omitted from the very beginning \cite{Gora97}.
Then, using Eq.(\ref{cd1}) and treating the terms containing
$S^-$ and $S^+_{\pm}$ as perturbations, we obtain
$\varepsilon_{\nu}=\varepsilon_{\nu}^{(0)}+\varepsilon_{\nu}^{(1)}$,
where $\varepsilon_{\nu}^{(0)}$ is the excitation energy in the mean
field approach, and the correction to the excitation energy
$\varepsilon_{\nu}^{(1)}=\delta\varepsilon_{\nu}-i\Gamma_{\nu}$ is
given by the relation
\begin{equation}      \label{pertenergy}
\varepsilon_{\nu}^{(1)}\!=\!\langle f^-_{\nu}| S^-|
f^+_{\nu}\rangle +
\frac{1}{2}\left(\langle f^-_{\nu}| S^+_+ +2\beta
\sqrt{\tilde n_0}\!|f^-_{\nu}
\rangle +
\langle f^+_{\nu}| S^+_-|f^+_{\nu}\rangle
\right).
\end{equation}
Here $f_{\nu}^{\pm (0)}$ are the zero-order wavefunctions of the excitations,
determined by the ordinary Bogolyubov-De Gennes equations~(\ref{fplus0}),
(\ref{fmin0}), with $\varepsilon_{\nu}=\varepsilon_{\nu}^{(0)}$.

In the case of quasiclassical excitations also the kernels of
resonant parts of integral operators in Eq.(\ref{pertenergy}) vary on a
distance scale $|{\bf r}-{\bf r'}|$ which does not exceed the correlation
length $l_{\rm cor}$.
This can be already seen from Eqs.~(\ref{Q}), (\ref{K2}): The characteristic
time scale $1/\varepsilon_{\nu}$ in Eq.(\ref{Q}) is much shorter than
$\omega^{-1}$ and important is only a small part of the classical
trajectory, where the condensate density is practically constant and
${\bf r}_p(t)={\bf r}+{\bf v}t$, with ${\bf v}=\partial H/\partial
{\bf p}$.
The correlation length $l_{\rm cor}$ is not only much smaller than the
size of the condensate, but also smaller than the width of the boundary
region of the condensate, where $n_0\widetilde{U}\sim \varepsilon_{\nu}$.
Therefore, the action of all integral
operators on the functions $f_{\nu}^{\pm (0)}$ in Eq.(\ref{pertenergy})
can be calculated in the local density approximation.
Accordingly, for each of these operators one
can use the quantity following from Eqs.~(\ref{Sigman})-(\ref{sigmaar}),
with $n_0=\tilde n_0({\bf r})$ and $p$ from the Bogolyubov
dispersion law $H({\bf p},{\bf r})=\varepsilon_{\nu}$. Then, using Eqs.~(\ref{form})
we can express the
energy shift $\delta\varepsilon_{\nu}$ and the damping rate $\Gamma_{\nu}$
through the energy shift
$\delta\varepsilon_{\nu h}({\bf r})$ and damping rate
$\Gamma_{\nu h}({\bf r})$ of the excitation of energy $\varepsilon_{\nu}$
in a spatially homogeneous Bose-condensed gas with the condensate density
equal to $\tilde n_0({\bf r})$:
\begin{eqnarray}       \label{ecorqhom}
\!\!\!\!\delta\varepsilon_{\!\nu}\!\!\!&\!=\!\!&\!\!\!\int\!\!\!d^3\!r
|f_{\!\nu}({\bf r})|^2\!\!\left\{\!\!\delta\varepsilon_{\!\nu h}({\bf r})\!
+\!\!\frac{\varepsilon_{\nu}\beta\sqrt{\tilde n_0({\bf r})}}
{\sqrt{\!\varepsilon_{\nu}^2\!\!+\!(\tilde n_0({\bf r})\tilde U)^2}\!\!+\!
\tilde n_0({\bf r})\tilde U}\!\!\right\}\!\!,\!\!\! \label{ecorghoms} \\
\!\!\Gamma_{\nu}&=&\int d^3r|f_{\nu}({\bf r})|^2\Gamma_{\nu h}({\bf r}).
\label{ecorghomd}
\end{eqnarray}
The second term in the integrand of Eq.(\ref{ecorghoms}) originates from
the temperature dependence of the shape of the condensate wavefunction.
For any ratio $\varepsilon_{\nu}/n_0({\bf r})\widetilde{U}$ this positive
term dominates over the negative term $\delta\varepsilon_{\nu h}({\bf r})$.
The latter circumstance can be easily established from
the results for $\delta\varepsilon_{\nu h}$ in Fig.3.
Thus, for quasiclassical low-energy excitations the energy shift
$\delta\varepsilon_{\nu}$ will be always positive, irrespective of the
trapping geometry and the symmetry of the excitation.

We confine ourselves to the case of cylindrical symmetry, where for
the states with zero angular momentum one finds
\begin{equation}         \label{zerom}
|f_{\nu}({\bf r})|^2=\frac{\tilde\mu}{\pi l_{\rho}l_z
\log{(2\tilde \mu/\varepsilon_{\nu})}\rho\sqrt{\varepsilon_{\nu}^2+
(\tilde n_0({\bf r})\tilde U)^2}},
\end{equation}
with $l_{\rho}=(2\tilde\mu/\omega_{\rho})^{1/2}$,
$l_z=(2\tilde\mu/\omega_z)^{1/2}$ being the characteristic size of the
condensate in the radial and axial direction, $\omega_{\rho}$, $\omega_z$
the radial and axial frequencies, and $\rho$ the radial coordinate.
The main contribution to the integral in Eq.(\ref{ecorqhom}) comes from
the boundary region of the condensate, where $\tilde n_0({\bf r})\sim
\varepsilon_{\nu}$. From Eqs.~(\ref{fplus}),(\ref{fmin}) one can easily
see that in this region the possibility to omit the non-Thomas-Fermi effects
originating from the kinetic energy of the condensate requires the condition
$\varepsilon_{\nu}\gg\omega^{2/3}\tilde\mu^{1/3}$. This condition ensures
that the characteristic width of the boundary region greatly exceeds the
excitation wavelength, and we arrive at the following
relations for the energy shifts and damping rates of the excitations:
\begin{eqnarray}
\delta\varepsilon_{\nu}&\approx & 8\sqrt{\frac{\varepsilon_{\nu}}
{\tilde \mu}}
\frac{T}{\log{(2\tilde \mu/\varepsilon_{\nu})}}(\tilde n_{0m}a^3)^{1/2},
\label{qcshift} \\
\Gamma_{\nu}&\approx & 9\sqrt{\frac{\varepsilon_{\nu}}{\tilde \mu}}
\frac{T}{\log{(2\tilde \mu/\varepsilon_{\nu})}}(\tilde n_{0m}a^3)^{1/2}.
\label{qcdamp}
\end{eqnarray}

It is important to emphasize that in the boundary region of the condensate,
responsible for the energy shifts and damping rates of the quasiclassical
excitations, the quantities $S^{-(1)}$, $S^{+(1)}$, and $\Sigma_a^{(1)}$
are determined by the contribution of intermediate quasiparticles which
have energies comparable with $\varepsilon_{\nu}$. Moreover, in this
spatial region the quasiparticle energies are of order the local mean
field interparticle interaction. As a consequence, the energy shift
$\delta\varepsilon_{\nu}$ (\ref{qcshift}) and the damping rate
$\Gamma_{\nu}$ (\ref{qcdamp}) are practically independent of the condensate
density profile. For the same reason the damping rate is determined by both
the Szepfalusy-Kondor and Beliaev damping processes.
Therefore, similarly to the damping of excitations with energies $\varepsilon_{p}
\agt n_0\widetilde{U}$ in a spatially homogeneous gas, the damping of
quasiclassical low-energy excitations of a trapped Bose-condensed gas can no
longer be treated as Landau damping.

\section{Sound waves in cylindrical Bose condensates}

The derivation of Eqs.~(\ref{qcshift}), (\ref{qcdamp}) assumes that the
motion of the excitation $\nu$ is quasiclassical for all degrees of
freedom. We now turn to the condensate excitations
in cigar-shaped
cylindrical traps, which are quasiclassical only in the axial direction
and  correspond to the lowest modes of the radial motion.
We will consider low-energy excitations
($\varepsilon_{\nu}\ll n_{0m}\widetilde{U}$), i.e., the excitations with
the axial wavelength much larger than the correlation length $l_{\rm cor}$.
In the recent MIT experiment \cite{LastKetterle} localized excitations
of this type were created in the center of the trap by modifying the
trapping potential using the dipole force of a focused off-resonant
laser beam. Then, a wave packet traveling along the axis of the
cylindrical trap (axially propagating sound wave) was observed.
In the mean field approach the sound waves propagating in an
infinitely long (axially homogeneous) cylindrical Bose condensate
have been discussed in \cite{Zaremba,Kav,S}.

For revealing the key features of the non-mean-field effects (damping
and the change of the sound velocity) we confine
ourselves to the same trapping geometry.
With regard to realistic cylindrical traps this will be a good approach
if the mean free path of sound waves is smaller than
the characteristic axial size of the sample.
As found in \cite{Zaremba},
for axially propagating sound waves radial oscillations of the condensate
are absent, and the wavefunctions $f_k^{\pm}=(2\tilde n_{0}(\rho)
\widetilde{U}/\varepsilon_k)^{\pm 1/2}f_k$, with
\begin{equation}    \label{fk}
f_k=\frac{1}{\sqrt{\pi l_{\rho}^2}}\exp{(ikz)}
\end{equation}
and $k$ being the axial momentum.
The dispersion law
\begin{equation}     \label{disp}
\varepsilon_k=ck
\end{equation}
is characterized by the sound velocity equal to
$(\tilde n_{0m}\widetilde{U}/2)^{1/2}$, where
$\tilde n_{0m}=\tilde\mu/\widetilde{U}$ is the maximum density of the
Thomas-Fermi condensate in the ordinary mean field approach.

It should be noted from the very beginning that, according to Eq.(\ref{cd1}),
$\tilde n_{0m}$ is related to the corrected value of the
maximum condensate density $n_{0m}$ as $\tilde n_{0m}=
n_{0m}-(\beta/\widetilde{U})\sqrt{n_{0m}}$. Therefore, being interested
in the sound velocity at a given value of the maximum condensate density,
one should substitute this expression to Eq.(\ref{disp}). This immediately
changes the sound velocity to
\begin{equation}         \label{sound}
c=(n_{0m}\widetilde{U}/2)^{1/2}
\end{equation}
in the leading term (\ref{disp}) of the dispersion law and provides a
contribution to the frequency shift of the sound wave
\begin{equation}      \label{s1}
\delta\overline{\varepsilon}_k=-\varepsilon_k\frac{6T}
{n_{0m}\widetilde{U}}(\pi n_{0m}a^3)^{1/2}.
\end{equation}

The damping rate and other contributions to the frequency shift can be
found directly from
Eq.(\ref{pertenergy}) by using the wavefunctions $f_k$ (\ref{fk}).
The intermediate quasiparticles giving the main
contribution to the damping rate and frequency shift have energies
$\varepsilon\sim n_{0m}\widetilde{U}$, i.e. much larger than the frequency
of the considered sound wave, $\varepsilon_k$ (see below).
Therefore, similarly to the case of phonons in a spatially homogeneous
condensate, the non-resonant terms (\ref{TNSn}), (\ref{TNSa})
contribute only to the frequency shift. As already mentioned above, the
characteristic distance
scale $|\mbox{\boldmath $\rho$}-\mbox{\boldmath $\rho$}^{\prime}|$ in the
kernels of the self-energies (\ref{TNSn}), (\ref{TNSa})
is of order the correlation length $l_{\rm cor}$, and the sum of their
contributions to the frequency shift, $\delta\varepsilon_k^n$, can be
calculated by using the local density approximation for the action of the
self-energy operators on the functions $f_k^{\pm}$. As a result, we express
$\delta\varepsilon_k^n$ through the non-resonant part of the energy shift
$\delta_{kh}^n(\rho)$ in the spatially homogeneous condensate with the
condensate density $\tilde n_0(\rho)$:
$$
\delta\varepsilon_k^n=\int d^2\rho |f_k|^2\left\{\delta_{kh}^n(\rho)
+\frac{\beta\sqrt{\tilde n_0(\rho)}}{2\tilde n_0(\rho)
\widetilde{U}}\varepsilon_k\right\}.
$$
For $\varepsilon_k\ll n_{0}\widetilde U$ one can directly find from
Eqs.~(\ref{Sigman}), (\ref{sigmaan}) that $\delta_{kh}^n=
\varepsilon_k(T/n_0\widetilde{U})(\pi n_0a^3)^{1/2}$. Then we obtain
\begin{equation}\label{dnonres}
\delta\varepsilon_k^n=
\varepsilon_k\frac{5T}{n_{0m}\widetilde{U}}(\pi n_{0m}a^3)^{1/2}.
\end{equation}

The resonant terms (\ref{TRSn}), (\ref{TRSa}) contribute to both the
frequency shift and damping rate.
This means that the latter is determined by the Szepfalusy-Kondor
scattering processes and, since the characteristic energies of
intermediate quasiparticles are much larger than $\varepsilon_k$, can be
treated as Landau damping.
The resonant contributions to the frequency shift and damping rate can not
be found in the local density approximation, as the characteristic
distance scale $|\mbox{\boldmath $\rho$}-\mbox{\boldmath $\rho$}^{\prime}|$
in the kernels of the self-energy operators
in Eq.(\ref{pertenergy}) is of order the radial size of the condensate.
For finding these contributions one has to substitute the resonant parts of
the self-energies, (\ref{TNSn})-(\ref{TRSa}), to Eq.(\ref{pertenergy}) and,
by using Eqs.(\ref{Q})-(\ref{K2}), turn from summation over
quasiclassical states $\gamma$, $\gamma'$ of intermediate quasiparticles
to the integration along classical trajectories of their motion. Then, a
direct calculation yields (cf. \cite{FSW})
\begin{equation}          \label{deltak1}
\delta^r_k=i\frac{\varepsilon_k^2\tilde U}{2}
\int d\varepsilon_{\gamma}\frac{dn_{\gamma}}{d\varepsilon_{\gamma}}
\!\int_0^{\infty}\!\!\! dt\exp{(i\varepsilon_k t) }\!\int
\Phi_{k\gamma}({\bf r})\Phi^{*}_{k\gamma}({\bf r}({\bf p},t))
\delta(\varepsilon_{\gamma}\!-\!H({\bf p},{\bf r}))\frac{d^3rd^3p}
{(2\pi)^3},
\end{equation}
where ${\bf r}({\bf p},t)$ is the classical trajectory starting at the
phase space points $({\bf r},{\bf p})$ on the (hyper)surface of constant
energy $\varepsilon_{\gamma}$,
$\Phi_{k\gamma}({\bf r})=f_k(z)F_{\gamma}(\rho)$,
and
$$
F_{\gamma}(\rho)\!=\!\frac{2\varepsilon_{\gamma}^2\!+\!(\tilde n_0(\rho)
\tilde U)^2\!-\!\tilde n_0(\rho)\tilde U\!
\sqrt{\varepsilon_{\gamma}^2\!+\!(\tilde n_0(\rho)\tilde U)^2}}
{\varepsilon_{\gamma}\sqrt{\varepsilon_{\gamma}^2\!+\!(\tilde n_0(\rho)
\tilde U)^2}}.
$$

Generally speaking, the integration in Eq.(\ref{deltak1}) is a tedious
task as it requires a full knowledge of the classical trajectories on a
time scale $\sim 1/\varepsilon_k$.
This is also the case in the idealized cylindrical trap, because of coupling
between the radial and axial degrees of freedom.
We will rely on the approach which assumes a fast radial motion of
quasiparticles compared to their motion in the axial direction and, hence,
requires the frequency of the sound wave, $\varepsilon_k$, significantly
smaller than the radial frequency $\omega_{\rho}$.
Then on a time scale $\sim 1/\varepsilon_k$ the quasiparticles with
energies $\sim \tilde n_0\tilde U$ (which are the most important for the
energy shifts and damping of the sound wave) oscillate many times in the
radial direction, whereas their axial variables $z({\bf p},t)$,
$p_z({\bf p},t)$ only slightly
change and, hence, can be adiabatically separated from the fast radial
variables $\mbox{\boldmath $\rho$}({\bf p},t)$, ${\bf p}_{\rho}({\bf p},t)$.
In this case it is convenient to integrate Eq,(\ref{deltak1}) over
$d{\bf p}_{\rho}$  and, using Eq.(\ref{fk}), represent it in the form
\begin{equation}          \label{interm}
\delta^r_k=i\frac{\varepsilon_k^2\widetilde U}{4\pi^2 l_{\rho}^2}
\int\varepsilon_k d\varepsilon_{\gamma}\frac{dn_{\gamma}}
{d\varepsilon_{\gamma}}\int dt\exp{(i\varepsilon_k t)}\int
\rho d\rho dp_zdz
\frac{F_{\gamma}(\rho)F_{\gamma}(\rho({\bf p},t))}
{\sqrt{\varepsilon_{\gamma}^2+(\tilde n_0(\rho)\widetilde U)^2}}
\exp{\{i(z-z({\bf p},t))\}},
\end{equation}
where the integration is performed over the entire classically accessible
region of the phase space.

Since $z({\bf p},t)$ is close to $z$, in the exponent of the integrand we
can write $z({\bf p},t)-z=v_z$, where the axial velocity $v_z$ is obtained
from the exact Hamiltonian equations of motion by averaging over the fast
radial variables: $v_z=\langle \partial H({\bf p},{\bf r})/\partial p_z
\rangle_{\rho}$.
For the classical radial motion ($\varepsilon_{\gamma}\gg\omega_{\rho}$)
the averaging procedure simply reduces to integration over $d\rho$ under
the condition $H({\bf p},{\bf r})=\varepsilon_{\gamma}$ at fixed values of
$\varepsilon_{\gamma}$, $p_z$ and $z$, with the weight proportional to
the local density of states for the radial motion:
$$
\langle(...)\rangle_{\rho}=g^{-1}\int (...)(\varepsilon_{\gamma}^2+
(\tilde n_0(\rho)\widetilde{U})^2)^{-1/2}2\pi\rho d\rho,
$$
where $g=\int (\varepsilon_{\gamma}^2+(\tilde n_0(\rho)
\widetilde U)^2)^{-1/2}2\pi\rho d\rho$.
Finally, averaging the function $F_{\gamma}(\rho(t))$ over the fast radial
variables and integrating over $dt$ in Eq.(\ref{interm}), we obtain
\begin{equation}       \label{dkf}
\delta_k^r=\frac{\varepsilon_k^2\widetilde U}{8\pi^3l_{\rho}^2}\int
d\varepsilon_{\gamma}\frac{\varepsilon_{\gamma}dn_{\gamma}}
{d\varepsilon_{\gamma}}
\int dzdp_z g\frac{\langle F(\rho)\rangle^2_{\rho}}
{\varepsilon_k-p_zv_z+i0}.
\end{equation}
The resonant contribution to the frequency shift, given by the real part of
Eq.(\ref{dkf}), after the integration proves to be $\delta\varepsilon_k^r
\approx -2.3\varepsilon_k(T/n_{0m}\widetilde{U})(n_{0m}a^3)^{1/2}$.
The sum of this quantity with the non-resonant term (\ref{dnonres}) and
$\delta\overline{\varepsilon}_k$ (\ref{s1}) leads to the frequency shift
of the sound wave
\begin{equation}
\delta\varepsilon_k\approx -4\varepsilon_k(n_{0m}a^3)^{1/2}
\frac{T}{n_{0m}\widetilde{U}}.\label{sndshift}
\end{equation}
The imaginary part of Eq.(\ref{dkf}) gives the damping rate
\begin{equation}   \label{sd}
\Gamma_k=8.6\varepsilon_k(n_{0m}a^3)^{1/2}\frac{T}{n_{0m}\widetilde{U}}.
\end{equation}

Except for the numerical coefficients, Eqs,(\ref{sndshift}),
(\ref{sd}) are similar to Eq.(\ref{reshomRe}), (\ref{reshomIm}) for the
damping  rate and energy shift of phonons in a  spatially homogeneous Bose
condensate. This a consequence of the fact that the condensate boundary
region practically does not contribute to the damping rate and frequency
shift of axially propagating sound waves, in contrast to the case of
excitations quasiclassical for both axial and radial degrees of freedom.

In the MIT experiment \cite{LastKetterle} the characteristic spatial
size of created localized excitations was $\lambda\approx 20$ $\mu$m
and, accordingly, so was the initial wavelength of propagating sound.
According to the experimental data, the propagating pulse died out during
25 ms, and after that only the lowest quadrupole excitation characterized
by a much loner damping time ($\sim 300$ ms) was observed.
We believe that the attenuation of axially propagating sound in the MIT
experiment \cite{LastKetterle} on the time scale of
25 ms can be well explained as a consequence of damping. The characteristic
frequency of the waves in the packet can be
estimated as $\varepsilon_{\nu}\approx 2\pi \sqrt{n_{0m}\tilde U/2}/\lambda$.
Then Eq.(\ref{sd}) gives the damping rate independent of the condensate
density $\tilde n_{0m}$.
In the MIT experiment the temperature $T\approx$0.5 $\mu$K was roughly
only twice as large as $n_{0m}\widetilde{U}$ , which decreases the damping
rate by approximately $20\%$ compared to that given by Eq.(\ref{sd}). In
these conditions we obtain a characteristic damping time of $15$ ms,
relatively close to the measured value.

The relative change of the sound velocity, $\delta c/c=
\delta\varepsilon_k/\varepsilon_k$, increases with decreasing condensate
density $n_{0m}$. However, even at the lowest densities of the MIT
experiment \cite{LastKetterle} ($n_{0m}\approx 10^{14}$ cm$^{-3}$)
the quantity $\delta c/c$ does not exceed $\sim 5\%$ and is practically
invisible.

\section*{Acknowledgements}

We acknowledge fruitful discussions with
M.W. Reynolds, M.A. Baranov and A. Griffin.
This work was supported by the Dutch Foundation FOM and by the Russian
Foundation for Basic Studies.

\end{document}